\documentclass[]{spie}  

\usepackage{amsmath,amsfonts,amssymb}
\usepackage{graphicx}
\usepackage[colorlinks=true, allcolors=blue]{hyperref}
\usepackage{comment}

\usepackage{multirow}
\title{Unsupervised Anomaly Detection of Paranasal Anomalies in the Maxillary Sinus}

\author[a]{Debayan Bhattacharya}
\author[b]{Finn Behrendt}
\author[c]{Benjamin Tobias Becker}
\author[d]{Dirk Beyersdorff}
\author[e]{Elina Petersen}
\author[f]{Marvin Petersen}
\author[g]{Bastian Cheng}
\author[h]{Dennis Eggert}
\author[i]{Christian Betz}
\author[j]{Anna Sophie Hoffmann*}
\author[k]{Alexander Schlaefer*}
\affil[a,b,k]{Hamburg University of Technology, Hamburg, Germany}
\affil[a,c,d,e,f,g,h,i,j]{ Universitätsklinikum Hamburg-Eppendorf, Hamburg, Germany}

\authorinfo{Further correspondence, please send an email to: debayan.bhattacharya@tuhh.de\\
* These authors contributed equally.}

\begin{document} 
\maketitle

\begin{abstract}

Deep learning (DL) algorithms can be used to automate paranasal anomaly detection from Magnetic Resonance Imaging (MRI). However, previous works relied on supervised learning techniques to distinguish between normal and abnormal samples. This method limits the type of anomalies that can be classified as the anomalies need to be present in the training data. Further, many data points from normal and anomaly class are needed for the model to achieve satisfactory classification performance. However, experienced clinicians can segregate between normal samples (healthy maxillary sinus) and anomalous samples (anomalous maxillary sinus) after looking at a few normal samples. We mimic the clinicians ability by learning the distribution of healthy maxillary sinuses using a 3D convolutional auto-encoder (cAE) and its variant, a 3D variational autoencoder (VAE) architecture and evaluate cAE and VAE for this task. Concretely, we pose the paranasal anomaly detection as an unsupervised anomaly detection problem. Thereby, we are able to reduce the labelling effort of the clinicians as we only use healthy samples during training. Additionally, we can classify any type of anomaly that differs from the training distribution. We train our 3D cAE and VAE to learn a latent representation of healthy maxillary sinus volumes using L1 reconstruction loss. During inference, we use the reconstruction error to classify between normal and anomalous maxillary sinuses. We extract sub-volumes from larger head and neck MRIs  and analyse the effect of different fields of view on the detection performance. Finally, we report which anomalies are easiest and hardest to classify using our approach. Our results demonstrate the feasibility of unsupervised detection of paranasal anomalies from MRIs with an AUPRC of 85\% and 80\% for cAE and VAE, respectively.        

\end{abstract}

\keywords{paranasal anomaly, unsupervised anomaly detection, autoencoder, VAE}

\section{INTRODUCTION}
\label{sec:intro}  
Anomalies occurring in the paranasal sinuses are commonly reported in patients who undergo neuroradiological assessment of the head using diagnostic imaging \cite {Wilson.2017}. These incidental findings pose clinical challenges\cite{Hansen.2014} and we have limited knowledge on the importance of these reported findings on the general population. To this end, numerous studies have been done to analyse the significance of these findings \cite{Tarp.2000,Rak.1991,Stenner.2014,Rege.2012,Cooke.1991}. Most of these works are population studies involving large sample sizes which are manually annotated by clinicians. In a three year retrospective study, it was observed that malignant tumours and inverted papillomas were classified as nasal polyps with a misdiagnosis rate of 5.63\% and 8.45\% respectively. Therefore, computer aided diagnosis systems (CADx) \cite{Kim.2019,Jeon.2021,Liu.2022} have been proposed with the idea of working in conjunction with the clinician to reduce the misdiagnosis rate of paranasal anomalies. However, all these works rely on supervised learning and consider at most one anomaly. Apart from requiring large labelled datasets, supervised learning models also need labelled data that are representative of the classes for accurate prediction \cite{app11020796}. Further, the type of anomaly to be classified has to be decided beforehand. In our case, we consider three anomalies, namely: (i) mucosal thickening (ii) polyps (iii) cysts. This is particularly challenging in our case where the considered anomalies  are known to have high intra-class morphological variations \cite{Tos2000,Janner2011-ug,Hung2021-lx} and have unequal occurrences in our dataset.

In light of the aforementioned points, we are motivated to perform Unsupervsised Anomaly Detection (UAD) using autoencoders. Autoencoders are a common choice for data compresison and outlier detection \cite{10.1145/3439950}. In medical imaging, brain anomaly detection and segmentation \cite{baur2021autoencoders} has gained popularity over the years.  The underlying concept of reconstruction-based UAD is that the autoencoder learns to compress and reconstruct only normal images during training. The assumption is that during testing, reconstruction errors will be low for normal images whereas anomalous images will have a large reconstruction error as the autoencoder will fail to properly reconstruct anomalous regions. In our case, we use cAE and VAE learn to compress and reconstruct healthy maxillary sinus (MS) volumes. Through our approach, we derive multiple benefits, namely: (i) our autoencoder becomes indifferent to the anomaly distribution thereby allowing detection of more than one anomaly, (ii) we reduce the labelling effort of the clinicians as the training dataset does not require anomalous samples,  (iii) we are able to generate a heat map based on the reconstruction error between the original volume and the reconstructed volume. The heat maps visualize the region of the potential anomaly and highlight it. This may prove to be beneficial to the clinician when making a diagnosis. 

In summary, our contributions are three-fold. First, we pose the paranasal anomaly detection problem as a UAD problem and thereby become indifferent to the anomaly distribution.  Second, we systematically evaluate our cAE and VAE approach on MS volumes with different field of view. Third, we report which anomalies are easiest and the hardest to classify using our UAD method. 

\section{METHODS}
\subsection{Dataset and Implementation Details}
Our labelled dataset consists of head and neck MRIs of 199 patients. Each MRI is a fluid attenuated inversion recovery (FLAIR) MRI. Our labelled dataset is part of the Hamburg City Health Study \cite{Jagodzinski2020}. Out of the 199 patients, 93 patients exhibit one or multiple anomalies in at least left or right MS.  Two Ear, Nose and Throat (ENT) surgeons and one ENT specialised radiologist confirmed the diagnosis of the observed pathology. We group the 3 anomalies into a single class called "anomaly"  and the normal MS are categorized into "normal" class. Altogether, we have 269 normal MS volumes and 130 anomalous MS volumes .  Each MRI has a resolution of 173 x 319 x 319 voxels along the saggital, coronal and axial directions respectively with each voxel of size 0.53 \(mm\) x 0.75 \(mm\) x 0.75 \(mm\). 

\textbf{Preprocessing} : We randomly selected a FLAIR MRI as the fixed MRI and performed rigid registration on the remaining MRIs. This was followed by resampling to a dimension of 128 \(\times\) 128 \(\times\) 128. We extracted two sub-volumes from the resampled head and neck MRIs, one for each MS. The extracted sub-volumes were of sizes: 33 \(\times\) 47 \(\times\) 45 (small), 46 \(\times\) 57 \(\times\) 55 (medium) and 53 \(\times\) 67 \(\times\)65 (large).  Owing to the symmetry of left and right MS, we horizontally flipped the coronal planes of right MS to give it the appearance of left MS for each patient. Finally, these sub-volumes were reshaped to a standard size of 64 \(\times\) 64 \(\times\) 64 voxels for the 3D cAE and VAE. All the MS volumes were normalised to a range of 0 to 1. Our preprocessing pipeline is illustrated in figure \ref{fig1} (a,b)

\textbf{Data split}: Our training set contains 172 normal MS volumes , validation set contains 43 normal MS volumes and 52 anomalous MS volumes and test set contains 54 normal MS volumes and 78 anomalous MS volumes. We perform a three-fold cross validation split for all our experiments.  

\textbf{Implementation Details}: We use PyTorch \cite{Paszke.03.12.2019} and PyTorch Lightning \cite{falcon2019pytorch} for all our experiments. We use batch size \(N\) = 16 and latent dimension \(n_{z}\) = 512 for the cAE. We use a learning rate of \(1e^{-4}\) and Adam optimizer \cite{https://doi.org/10.48550/arxiv.1412.6980} with default parameters to train our cAE and VAE. We run all our experiments for 100 epochs.  

\subsection{Deep Learning Methods}

\begin{figure}[hbt!]
\centering
\includegraphics[width=0.75\textwidth]{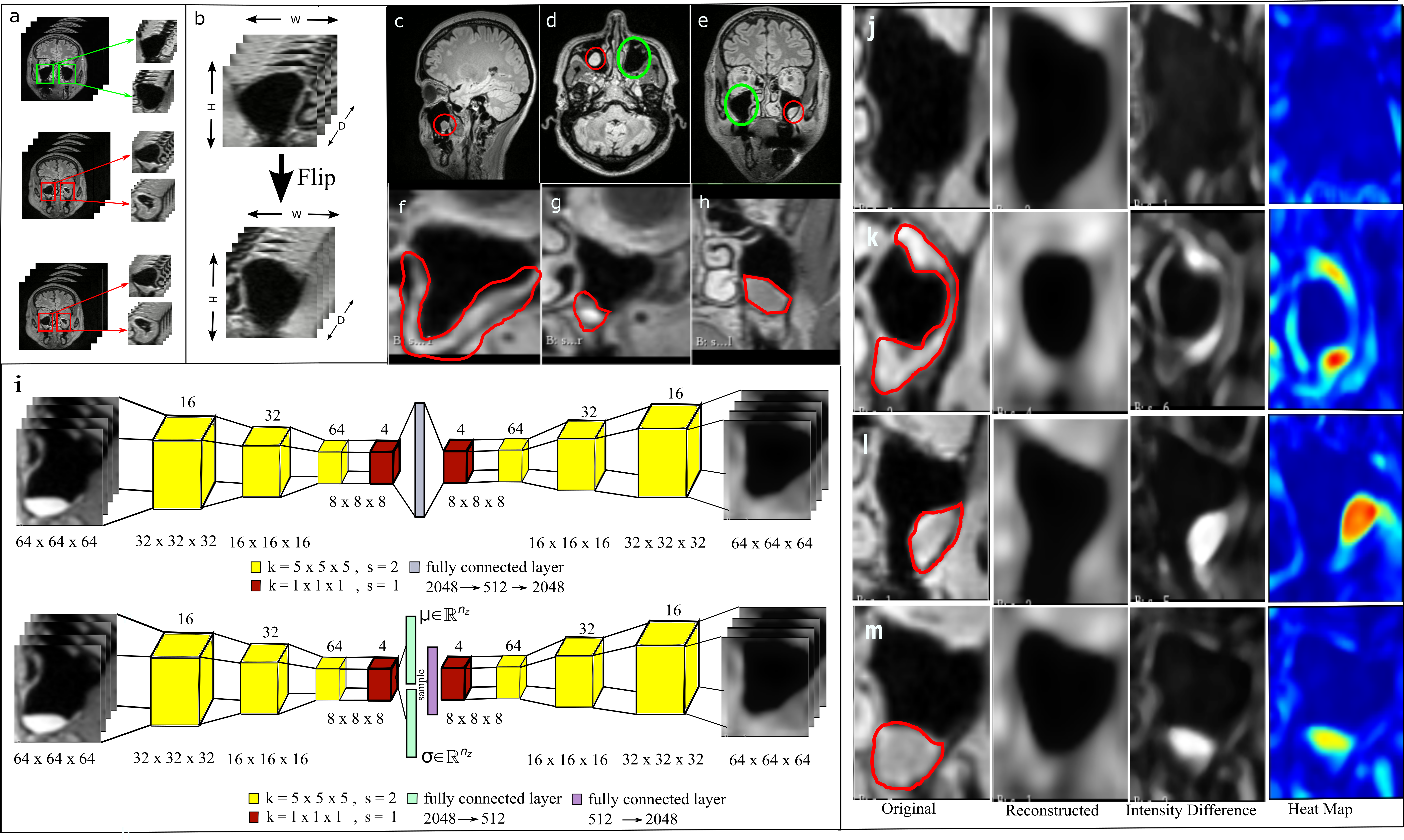}
\caption{(a) Extraction of left and right MS from head and neck MRI (b) Flipping of the coronal plane of right MS (c) Cyst in the right MS (d) Polyp in the left MS (e) Cyst in the left MS (f) MS (small) showing mucosal thickening (g) MS (medium) showing polyp (h) MS (large) showing cyst (i) TOP: Our cAE architecture with latent vector of size 512 used as bottleneck. BOTTOM: Our VAE architecture. In both the cAE and VAE decoders, we perform 3D convolution followed by trilinear upsampling.  ReLU is used as non-linear activation function in both the networks. (j)-(m) Images extracted from original , reconstructed , intensity difference and heat map volumes. (j) is an image extracted from normal MS volume. (k),(l),(m)  are mucosal thickening, polyps and cyst anomalies respectively. The red markings denote anomalies and green circles denote normal MS.}  \label{fig1}
\end{figure}
 Similar to Bengs \textit{et al} \cite{bengs2021three}, we extend our architecture to 3D as it has shown to improve the detection performance in 3D MRI scans. Our cAE and VAE architectures are shown in figure \ref{fig1} (i) . There are two stages to our approach. First, we train our cAE and VAE to learn the distribution of  \(X_{h}\) where \(X_{h}\) represents healthy MS volumes. For training our cAE, we use the L1 reconstruction loss as described below: 

\begin{equation}
    L_{1} = \sum_{k=1}^{N} |x^{k} - \hat{x}^{k}| 
\end{equation}

Here, \(x^{k}\) and \(\hat{x}^{k}\) denote the \(k\)-th MS volume and reconstructed MS volume respectively. N denotes the mini-batch size. For training our VAE, we use L1 reconstruction loss and KL Divergence. While training, the VAE learns a mean \(\mu_{z}\) and variance \(\sigma_{z}\) from which a sample is drawn and reconstructed. The loss used to train our VAE is shown below: 
\begin{equation}
    L_{VAE} = L_{1} + \lambda_{KL} D_{KL}(q(z|x)||p(z) )
\end{equation}

Here,  \(D_{KL}(.||.)\) represents the Kullback–Leibler divergence between the parameterized latent distribution \(q(z|x)\)  $\sim$ \(N ( \mu_{z},  \sigma_{z} )\) and the prior p(z) which
follows a multivariate normal distribution. \(z \in \mathbb{R}^{n_{z}}\) represents the latent vector. \(\lambda_{KL}\) is a Lagrangian multiplier and we have set it to 1 for our experiments. VAE projects the the input MS volume to \(q(z|x)\) and KL-Divergence loss attempts to bring it close to a prior \(p(z)\) \cite{https://doi.org/10.48550/arxiv.1312.6114}. 

Second, we use the trained cAE and VAE to reconstruct the MS volumes in the validation set. For each MS volume, we calculate the L1 and L2 reconstruction loss denoted as \(t_{L1}\) and \(t_{L2}\) respectively. We choose the optimal thresholds by plotting the precision recall curve and select the threshold with the highest F1 score. During inference, \(\hat{x}^{k}\) with \(L_{1} >\) \(t_{L1}\) and \(L_{2} >\)  \(t_{L2}\) is classified as anomalous MS volume. Here, \(L_{2}\) is the L2 reconstruction loss defined as follows: 

\begin{equation}
    L_{2} = \sum_{k=1}^{N} (x^{k} - \hat{x}^{k})^{2}
\end{equation} 

Additionally, for the MS volumes classified as anomalous, a further analysis is done by calculating voxel-wise intensity difference \(D_{k} = |x^{k} - \hat{x}^{k}| \) after which a median filter of kernel size 5 is applied on it to remove sporadic reconstruction errors. Finally, a heat map is rendered for the individual slices along the coronal, axial and sagittal planes as shown in figure \ref{fig1} (j, k, l, m). The regions in red denote the regions which the cAE failed to reconstruct. Our qualitative results indicate an overlap between the anomalous regions and the poorly reconstructed regions of MS volumes. 

\section{Results}

\begin{table}
\caption{Anomaly Detection Performance on two thresholds \(t_{L1}\) and \(t_{L2}\) where positive labels are assigned to the anomalous class. }
\centering
\resizebox{1.0\textwidth}{!}{

  \begin{tabular}{|l|l|l|l|l|l|l|l|l|l|}
    \hline
    Method &
    MS Size &
      \multicolumn{2}{c|}{Precision} &
      \multicolumn{2}{c|}{Recall} &
      \multicolumn{2}{c|}{F1}&
      \multicolumn{2}{c|}{AUPRC}\\
      
    & & $t_{L1}$ & $t_{L2}$ &  $t_{L1}$ & $t_{L2}$ &  $t_{L1}$ & $t_{L2}$ &  $t_{L1}$ & $t_{L2}$\\
    \hline
     VAE & small & 0.69 & 0.76 & 0.63 & 0.62 & 0.64 & 0.68 &0.76 & 0.80\\
    \hline
     VAE & medium & 0.63  & 0.63 & 0.84 & 0.91 & 0.72  & 0.75 & 0.70 & 0.75 \\
    \hline
     VAE & large & 0.61  & 0.64  & 0.81 & 0.86 & 0.70 & 0.73 & 0.65 & 0.69 \\
    \hline
     cAE & small & 0.75 & 0.81 & 0.74 & 0.66 & 0.74 & 0.73 & 0.81 & 0.83 \\
    \hline
    cAE & medium & 0.73 & 0.77 & 0.62 & 0.74 & 0.67 & 0.75 & 0.80 & 0.85 \\
    \hline
    cAE & large & 0.68 & 0.74 & 0.82 & 0.73 & 0.74 & 0.73 & 0.73 & 0.78 \\
    \hline
  \end{tabular}
  }
   \label{tab1}
\end{table}

\begin{table}[h!]
\caption{Accuracy  per anomaly on the test set reported in percentage (\%). Here, \# refers to the number of correctly classified samples divided by the total samples from a particular category.}
\centering
\resizebox{1.0\textwidth}{!}{

 \begin{tabular}{||c c c c c c ||} 
 \hline
 Method & MS Size & Normal (\%/\#) & Mucosal Thickening (\%/\#) & Polyps (\%/\#) & Cysts (\%/\#)  \\
 \hline\hline
 VAE & small & 0.48 (26/54) & 0.75 (22/29) &  0.82 (28/34) & 0.73 (11/15) \\ 
 cAE & medium & 0.61 (33/54) & 0.62 (18/29) &  0.91 (31/34) & 0.8 (12/15) \\

 \hline 
 \end{tabular}
 } \label{tab2}
\end{table}

The anomaly detection performance is shown in table \ref{tab1}. We consider the Area Under Precision Recall Curve (AUPRC) to evaluate our classifiers as we have an unbalanced test set. We rank the classifiers based on AUPRC. We report the mean values of the mentioned metrics. In terms of AUPRC, the VAE that uses small MS volume and cAE that uses medium MS volume are the best performing classifiers. 

From table \ref{tab1}, we notice that the all the VAEs have relatively lower AUPRC in comparison to cAEs. Furthermore, our results show that considering L2 loss when computing the anomaly score and using \(t_{L2}\) as threshold leads to better performance for all cAEs and VAEs. In table \ref{tab2}, we report the accuracy per anomaly in terms of percentage and number of occurances for the best performing VAE and cAE from table \ref{tab1}. We observe that polyps are the easiest to classify, followed by cysts. Mucosal thickening anomaly is the hardest to classify for both VAE and cAE.

\section{Discussion and Conclusion} 

From the results in table \ref{tab1} we observe that using \(t_{L2}\) as threshold leads to better performance. This can be attributed to the fact that L2 loss penalizes voxel-wise intesity outliers more heavily than L1 loss. Therefore, even small regions of poor reconstruction in the MS volume can amount to high overall reconstruction error. Additionally, our accuracy percentage per anomaly reported in table \ref{tab2} shows that mucosal thickening anomalies are the most difficult to classify. We believe this to be the case because unlike polyps and cysts which occur as visible masses in the MRI (See figure \ref{fig1} (l),(m)), mucosal thickening anomalies have more subtle appearances as these anomalies mostly cause inflammation of the mucosal walls. Therefore, unless the inflammations are too noticeable, they almost have the appearance of a healthy MS. We also observe that classifiers using small and medium MS volumes have the best AUPRCs. This can be attributed to the fact that cropping large volumes lead to inclusion of unnecessary surrounding anatomical structures outside of the MS. These additional  anatomical structures effect the reconstruction error and thereby, we end up selecting  sub-optimal thresholds.  Our work has some limitations, one being the accuracy of  healthy MS volume detection needs to be higher. We think this can be achieved by better localisation and cropping strategies of the MS in the head and neck MRI and by labelling more healthy MS volumes. Second, we have not experimented with autoencoders with skip connections or with more parameters and studied its effect on reconstruction error. 

In conclusion, we evaluate UAD for paranasal anomaly detection. Previous methods \cite{Kim.2019,Jeon.2021,Liu.2022} have used supervised learning methods and as a result are constrained to classify the anomalies that are included in the training distribution. Through our UAD approach, our models learn the healthy MS volume distribution \(X_{h}\) thereby reducing the labelling effort of the clinicians. Also, we are able to detect multiple anomalies. Further, we render heat maps of poor reconstruction. This can provide valuable insights to clinicians while making a diagnosis.

\bibliography{report} 
\bibliographystyle{spiebib} 

\end{document}